\documentclass[%
 pra,
 reprint,
 twocolumn,
 notitlepage,
 amsmath,amssymb,
 aps,
]{revtex4-1}
\usepackage{natbib}
\usepackage{graphicx,mathtools,amsmath,amssymb}
\usepackage[letterpaper,margin=2cm]{geometry}
\usepackage{dcolumn}
\usepackage{bm}
\usepackage{lipsum}
\usepackage{color}
\usepackage{mathtools}
\usepackage{xparse}
\NewDocumentCommand{\ceil}{s O{} m}{%
  \IfBooleanTF{#1} 
    {\left\lceil#3\right\rceil} 
    {#2\lceil#3#2\rceil} 
}

\newcommand*{\QED}{\hfill\ensuremath{\blacksquare}}

\begin{document}
\title{Bounds on Entanglement Catalysts}
\author{Michael Grabowecky}
\author{Gilad Gour}
\affiliation{Department of Mathematics and Statistics, and Institute for Quantum Science and Technology, University of Calgary, 2500 University Drive NW, Calgary, Alberta, Canada T2N 1N4}
\date{\today}
\begin{abstract}
Given a finite dimensional pure state transformation restricted by entanglement assisted local operations and classical communication (ELOCC), we derive minimum and maximum bounds on the entanglement of an ancillary catalyst that allows that transformation. These bounds are non-trivial even when the Schmidt number of both the original and ancillary states becomes large. We identify a lower bound for the dimension of a catalyst allowing a particular ELOCC transformation. Along with these bounds, we present further constraints on ELOCC transformations by identifying restrictions on the Schmidt coefficients of the target state. In addition, an example showing the existence of qubit ELOCC transformations with multiple ranges of potential ancillary states is provided. This example reveals some additional difficulty in finding strict bounds on ELOCC transformations, even in the qubit case. Finally, a comparison of the bounds in this paper with previously discovered bounds is presented. 
\end{abstract}
\maketitle
 \section{Introduction}
In recent years, entanglement has been identified as a valuable resource that has proven to be integral for use in multi-party quantum information protocols. Formally, entanglement can be defined as the resource which allows parties to overcome the limitations imposed by local operations and classical communication (LOCC) \cite{Nielsen2010, Jonathan1999}. Often entanglement is not available in its most pure form. Rather, a quantum system may be partially entangled and given in a form that is mixed with noise. Consequently, such systems may not be optimal for certain quantum information tasks. One can conclude that certain quantum states are more desirable than others depending on the objective at hand. For example, one may desire a Bell state for the most efficient operation of a quantum teleportation protocol. One important effect of LOCC is the conversion from one bipartite pure state to another. Such transformations involve the consumption of shared entanglement between parties such that the output system is less entangled than the input system. The precise characterization of these types of transformations is stated in Nielsen's Theorem~\cite{Nielsen1999}. 

Nielsen's Theorem is deeply rooted in the theory of majorization. Suppose that $\mathbf{p}$ and $\mathbf{q}$ are two vectors in $\mathbb{R}^d$. Then $\mathbf{p}$ is said to be \textit{majorized} by $\mathbf{q}$, denoted $\mathbf{p} \prec \mathbf{q}$, if:
\begin{equation}
\label{eq1}
\sum_{i = 1}^{l} p_{i}^{\downarrow} \leq \sum_{i = 1}^{l} q_{i}^{\downarrow} \qquad{       }\qquad \forall l \in \{1,2, \cdots ,d\}
\end{equation}
with equality when $l = d$. Here, $p_i^{\downarrow}$ is the $i$-th element of the vector $\mathbf{p}$ arranged in non-increasing order \cite{Diggle2010}. Every pure bipartite state has a Schmidt form ${|\psi^{AB} \rangle =  \sum_{i=1}^{d} \sqrt{p_i} |i^A \rangle |i^B \rangle}\in \mathcal{H}^{AB}$, where $p_i$ are known as the Schmidt coefficients of the state and have the property $p_i \geq 0 \ \forall \ i \in \{1,2, \cdots d \}$, ${\sum_{i=1}^{d}p_i = 1}$. Nielsen's Theorem states that if two parties share a pure state ${|\psi^{AB} \rangle =  \sum_{i=1}^{d} \sqrt{p_i} |i^A \rangle |i^B \rangle}\in \mathcal{H}^{AB}$, that they wish to transform into a second pure state ${|\phi^{AB} \rangle =  \sum_{j=1}^{d} \sqrt{q_{j}} |j^{A} \rangle |j^{B} \rangle \in \mathcal{H}^{AB}}$ using only LOCC, then the transformation $|\psi^{AB} \rangle \rightarrow |\phi^{AB} \rangle$ is possible with certainty if and only if $\mathbf{p} \prec \mathbf{q}$. However, a transformation from one state to another with unit probability is not always possible. Two bipartite states are said to be \textit{incomparable} if $|\psi \rangle \nrightarrow |\phi \rangle$ and $|\phi \rangle \nrightarrow |\psi \rangle$ by LOCC alone. To reconcile this problem, a shared ancillary state $|\chi^{A'B'} \rangle = \sum_{x=1}^{k} \sqrt{r_x} |x^{A'} \rangle |x^{B'} \rangle \in \mathcal{H}^{A'B'}$ can be borrowed to realize the desired transformation by increasing the entanglement of the initial state. Stated formally, $|\psi \rangle \nrightarrow |\phi \rangle$ and $|\phi \rangle \nrightarrow |\psi \rangle$, but $|\psi \rangle \otimes |\chi \rangle \rightarrow |\phi \rangle \otimes |\chi \rangle$ or equivalently, $\mathbf{p} \nprec \mathbf{q}$, $\mathbf{q} \nprec \mathbf{p}$, but $\mathbf{p} \otimes \mathbf{r} \prec \mathbf{q} \otimes \mathbf{r}$. This ancillary state can be borrowed as long as it is returned unchanged after the transformation has concluded. Thus, the two parties can obtain a previously unattainable state without using any extra entanglement. LOCC transformations performed in the presence of an ancillary (catalyst) state are called \textit{entanglement assisted} LOCC (ELOCC) transformations \cite{Jonathan1999}. 

The trouble with ELOCC transformations is that there is still no way to fully classify the properties of ancillary entangled states. Neccessary and sufficient conditions have recently been outlined on the existance of a catalyst state for any given transformation using the Renyi entropies and power means  \cite{Turgut2007, Klimesh2007}. However, these results did not present any conditions on the catalyst state itself, only on the existence of such a state. Furthermore, conditions found on catalyst states including the catalyst dimension have been presented in \cite{Sanders2009}. Yet, these conditions only work for a subset of catalytic examples. It is desirable to obtain a thorough understanding of ELOCC transformations as they play a vital role in quantum thermodynamics where a catalyst can be thought of as a heat engine undergoing a cyclic process that is returned to its original form when the thermal process has concluded \cite{Brandao}. In the case of \textit{thermomajorization} \cite{Horodecki2011}, it is athermality that is the resource rather than entanglement, however the underlying principles governing thermal state transformations remain consistent with ELOCC transformations \cite{Gouragain}.  

It has also been shown that any entangled target state can be \textit{embezzled} (up to a small amount of error) using only a family of bipartite catalysts  \cite{Hayden2002}. Unfortunately, embezzling with small error requires the catalyst Schmidt number to diverge to infinity which limits its utility. This paper will focus on finite dimensional ELOCC transformations in which embezzling is not possible with a high degree of accuracy. Specifically, bounds limiting the amount of entanglement a potential catalyst state can contain for any pure state transformation will be shown. These bounds restrict the values of specific Schmidt coefficients of the catalyst state and depend on Schmidt coefficients of the initial and target states. The bounds on catalyst entanglement are supplemented with additional conditions on the target state that must be satisfied for an ELOCC transformation to be possible. A lower bound for the dimension of a catalyst allowing a particular ELOCC transformation is also identified. Such a bound is somewhat of a rarity in the literature. In addition, the existence of transformations with multiple regions of qubit catalysis will be shown, which impose challenges for finding further bounds on ELOCC transformations, even in the qubit case. Finally, a brief comparison between the new bounds proved in this paper with the bounds presented in \cite{Sanders2009} will be shown.
\vspace{-0.25cm}
\section{\label{bounds}Bounds on Entanglement Catalysts} 
There are three important properties of ELOCC transformations that will be used extensively in this paper. All three properties were originally proved in \cite{Jonathan1999}. For the remainder of the paper, we assume that $\mathbf{p}$, $\mathbf{q} \in \mathbb{R}^d$ and $\mathbf{r} \in \mathbb{R}^k$ are all arranged in non-increasing order and that we are transforming from a state with Schmidt vector $\mathbf{p}$ to a state with Schmidt vector $\mathbf{q}$.
\newline
\newline
\textbf{Property 1:} For two incomparable Schmidt vectors $\mathbf{p}$, $\mathbf{q} \in \mathbb{R}^d$ and some catalyst vector $\mathbf{r} \in \mathbb{R}^k$ such that $\mathbf{p} \otimes \mathbf{r} \prec \mathbf{q} \otimes \mathbf{r}$, the largest element of $\mathbf{p}$ is always smaller than the largest element of $\mathbf{q}$ ($p_1 \leq q_1$) and the smallest element of $\mathbf{p}$ is always larger than the smallest element of $\mathbf{q}$ ($p_d \geq q_d$). In addition,
\begin{equation}
\label{eq4}
\sum_{i=1}^{d-1}p_i \leq \sum_{i=1}^{d-1} q_i
\end{equation}
\newline
\textbf{Property 2:} When $d = 2$, either $\mathbf{p} \prec \mathbf{q}$ or $\mathbf{q} \prec \mathbf{p}$, which makes borrowing a catalyst irrelevant. If $\mathbf{p}$, $\mathbf{q} \in \mathbb{R}^d$ are incomparable Schmidt vectors and $d = 3$, then no catalyst vector $\mathbf{r}$ exists such that $\mathbf{p} \otimes \mathbf{r} \prec \mathbf{q} \otimes \mathbf{r}$. 
\newline
\newline
\textbf{Property 3:} No transformation can be catalysed by the maximally entangled state, $\mathbf{r} = (1/k, 1/k, \cdots, 1/k)$.

\subsection{Bounds on Minimum and Maximum Entanglement of Catalyst States}
This section outlines bounds that quantify the amount of entanglement a catalyst state must contain in order to catalyse an incomparable state transformation.

Let $\mathbf{p}$ and $\mathbf{q}$ be incomparable Schmidt vectors. Then there exists at least one $l\in \{1,2,\cdots,d\}$ such that $\sum_{i = 1}^{l} p_{i} > \sum_{i = 1}^{l} q_{i}$, thus violating the majorization criterion (\ref{eq1}). Define the set of all $l$ values such that $\sum_{i = 1}^{l} p_{i} > \sum_{i = 1}^{l} q_{i}$ to be $\mathcal{L}$:
\begin{equation}
\label{set}
\mathcal{L} \equiv \left\{l\in\{1,2\cdots d\}\quad\bigg|\quad\sum_{i = 1}^{l} p_{i}-q_{i} > 0\right\}
\end{equation}
Let $m \equiv \min(\mathcal{L})$ and $n \equiv \max(\mathcal{L})$. Due to property 1, $m, \ n \neq 1, d-1$. Additionally, $m, \ n \neq d$ since ${\sum_{i=1}^{d}p_i = \sum_{i=1}^{d} q_i = 1}$. Because $p_1 \leq q_1$, $m$ can be thought of as the minimum $l$ causing $\mathbf{p}$ and $\mathbf{q}$ to be incomparable, while $n$ can be thought of as the maximum $l$ causing $\mathbf{p}$ and $\mathbf{q}$ to be incomparable.

Suppose that the Schmidt vector $\mathbf{r}$ is the product state $(1,\,0,\cdots,\,0)$. Then $\mathbf{r}$ cannot catalyse the transformation because the non-zero elements of $\mathbf{p} \otimes \mathbf{r}$, and  $\mathbf{q} \otimes \mathbf{r}$ are identical to $\mathbf{p}$ and $\mathbf{q}$ respectively. Similarly, if $\mathbf{r}$ is the maximally entangled state $(1/k, 1/k, \cdots, 1/k)$, then by property 3, $\mathbf{r}$ cannot catalyse the transformation. The question that arises is: How far can one deviate $\mathbf{r}$ from the product state or the maximally entangled state before an ELOCC transformation becomes possible? This question is resolved in Theorem 1. A related theorem was proven in \cite{Duan}, where bounds on single-copy ELOCC catalysts were extended to produce bounds on multi-copy ELOCC catalysts. Here we present bounds on single-copy ELOCC catalysts that are stronger than those presented in \cite{Duan}, but do not have application to multi-copy ELOCC catalysts.
\newline
\newline
\textbf{Theorem 1:}  For any incomparable Schmidt vectors $\mathbf{p}$, $\mathbf{q} \in \mathbb{R}^d$ and for any Schmidt vector $\mathbf{r} \in \mathbb{R}^k$, if $\mathbf{p} \otimes \mathbf{r} \prec \mathbf{q} \otimes \mathbf{r}$ , then $\mathbf{r}$ must satisfy:
\begin{equation}
\label{bound2}
\max_{v \in \{1,2,\cdots,k-1\}} \left(\frac{r_v}{r_{v+1}}\right) < \min\left(\frac{q_1}{q_{m}},\frac{q_{n+1}}{q_d}\right)
\end{equation}
\begin{equation}
\label{eqn5}
 \frac{r_1}{r_k} > \max_{l \in \mathcal{L}}\left(\frac{q_{l}}{q_{l+1}}\right)
\end{equation}
\textbf{Remark:} It appears that both bounds (\ref{bound2}) and (\ref{eqn5}) are only determined by the Schmidt vector $\mathbf{q}$. By definition, $q_l \ \forall \ l \in \mathcal{L}$ including $q_m$ and $q_{n}$ are identified using both the Schmidt vectors $\mathbf{p}$ and $\mathbf{q}$. Thus, (\ref{bound2}) and (\ref{eqn5}) are not independent of $\mathbf{p}$.
\newline
\newline
\textbf{Proof:} Let $\mathbf{p}$ and $\mathbf{q}$ be incomparable Schmidt vectors and let $\mathbf{p} \otimes \mathbf{r} \prec \mathbf{q} \otimes \mathbf{r}$ for some catalyst Schmidt vector $\mathbf{r}$. To begin, we prove that:
\begin{equation}
\label{pqrel}
\frac{p_1}{p_m} < \frac{q_1}{q_m}
\end{equation}
From the definition of $m$, we get:
\begin{equation}
\sum_{i=1}^{m}p_i > \sum_{i=1}^{m}q_i \qquad \text{and} \qquad \sum_{i=1}^{m-1} p_i \leq \sum_{i=1}^{m-1} q_i
\end{equation}
Therefore:
\begin{equation}
\sum_{i=1}^{m} p_i = \sum_{i=1}^{m-1} p_i + p_m > \sum_{i=1}^{m-1} q_i + q_m = \sum_{i=1}^{m} q_i
\end{equation}
This implies $p_m > q_m$. Because $\mathbf{p} \otimes \mathbf{r} \prec \mathbf{q} \otimes \mathbf{r}$, it holds that $p_1 \leq q_1$ by property 1. Combining $p_1 \leq q_1$ and $p_m > q_m$, we get:
\begin{equation*}
\frac{p_1}{p_m} < \frac{q_1}{q_m}
\end{equation*}
By a similar approach, we will now prove that:
\begin{equation}
\label{pqrel2}
\frac{p_{n+1}}{p_d} < \frac{q_{n+1}}{q_d}
\end{equation}
From the definition of $n$, we get:
\begin{equation}
\sum_{i=1}^{n}p_i > \sum_{i=1}^{n}q_i \qquad \text{and} \qquad \sum_{i=1}^{n+1} p_i \leq \sum_{i=1}^{n+1} q_i
\end{equation}
Therefore:
\begin{equation}
\sum_{i=1}^{n} p_i = \sum_{i=1}^{n+1} p_i - p_{n+1} > \sum_{i=1}^{n+1} q_i - q_{n+1} = \sum_{i=1}^{n} q_i
\end{equation}
This implies $p_{n+1} < q_{n+1}$. Because $\mathbf{p} \otimes \mathbf{r} \prec \mathbf{q} \otimes \mathbf{r}$, it holds that $p_d \geq q_d$ by property 1. Combining $p_d \geq q_d$ and $p_{n+1} < q_{n+1}$, we get:
\begin{equation*}
\frac{p_{n+1}}{p_d} < \frac{q_{n+1}}{q_d}
\end{equation*}
We are now ready to prove bound (\ref{bound2}). We use proof by contradiction. First, let:
\begin{equation}
\label{assume}
\frac{r_{v'}}{r_{v'+1}} \equiv \max_{v \in \{1,2,\cdots,k-1\}}\left(\frac{r_v}{r_{v+1}}\right) \geq \frac{q_1}{q_m}
\end{equation} 
Where $v'$ is the specific $v$ satisfying the maximum. Thus, $q_m r_{v'} \geq q_1 r_{v'+1}$. This implies that the first $(v'-1)d+m$ elements of $(\mathbf{q}\otimes \mathbf{r})^{\downarrow}$ consist of ${q_xr_y \ \forall \ x \in \{1,2,\cdots,d\}}$ and $\forall \ y \in \{1,2,\cdots,v'-1\}$ along with $q_{x'}r_{v'} \ \forall \ x' \in \{1,2,\cdots, m\}$. From (\ref{pqrel}) and (\ref{assume}), we get:
\begin{equation}
\frac{r_{v'}}{r_{v'+1}} \geq \frac{q_1}{q_m} > \frac{p_1}{p_m}
\end{equation}
Thus, the first $(v'-1)d+m$ elements of $(\mathbf{p}\otimes \mathbf{r})^{\downarrow}$ are analogous to $(\mathbf{q}\otimes \mathbf{r})^{\downarrow}$ and we get the following:
\begin{equation}
\begin{split}
\label{res1}
\sum_{i=1}^{(v'-1)d+m} (\mathbf{p} \otimes \mathbf{r})_i^{\downarrow} &= \sum_{j=1}^{v'-1}r_j + r_{v'}\sum_{h=1}^{m}p_h \\ 
& > \sum_{j=1}^{v'-1}r_j+r_{v'}\sum_{h=1}^{m}q_h \\
&= \sum_{i=1}^{(v'-1)d+m} (\mathbf{q} \otimes \mathbf{r})_i^{\downarrow}
\end{split}
\end{equation}
Where we used the fact that $\sum_{h=1}^{m}p_h > \sum_{h=1}^{m}q_h$ from the definition of $m$ and that $\sum_{h=1}^{d}p_h = \sum_{h=1}^{d}q_h = 1$. Note that the largest elements of $(\mathbf{p} \otimes \mathbf{r})^{\downarrow}$ and ${(\mathbf{q} \otimes \mathbf{r})^{\downarrow}}$ are always $p_1r_1$ and $q_1r_1$ respectively. From (\ref{res1}) and because $p_1 \leq q_1$ implies $p_1r_1 \leq q_1r_1$,  we conclude ${\mathbf{p} \otimes \mathbf{r} \nprec \mathbf{q} \otimes \mathbf{r}}$, which contradicts the assumption that ${\mathbf{p} \otimes \mathbf{r} \prec \mathbf{q} \otimes \mathbf{r}}$. Thus, we have proved that:
\begin{equation}
\label{halfbound}
\max_{v \in \{1,2,\cdots,k-1\}}\left(\frac{r_v}{r_{v+1}}\right) < \frac{q_1}{q_m}
\end{equation}
Next, let:
\begin{equation}
\label{assume2}
\frac{r_{v'}}{r_{v'+1}} \geq \frac{q_{n+1}}{q_d}
\end{equation}
Thus, $q_d r_{v'} \geq q_{n+1} r_{v'+1}$. This implies that the first $v'd+n$ elements of $(\mathbf{q}\otimes \mathbf{r})^{\downarrow}$ consist of ${q_xr_y \ \forall \ x \in \{1,2,\cdots,d\}}$ and $\forall \ y \in \{1,2,\cdots,v'\}$ along with $q_{x'}r_{v'+1} \ \forall \ x' \in \{1,2,\cdots, n\}$. From (\ref{pqrel2}) and (\ref{assume2}), we get:
\begin{equation}
\frac{r_{v'}}{r_{v'+1}} \geq \frac{q_{n+1}}{q_d} > \frac{p_{n+1}}{p_d}
\end{equation}
Thus, the first $v'd+n$ elements of $(\mathbf{p}\otimes \mathbf{r})^{\downarrow}$ are analogous to $(\mathbf{q}\otimes \mathbf{r})^{\downarrow}$ and we get the following:
\begin{equation}
\begin{split}
\label{res2}
\sum_{i=1}^{v'd+n} (\mathbf{p} \otimes \mathbf{r})_i^{\downarrow} &= \sum_{j=1}^{v'}r_j + r_{v'+1}\sum_{h=1}^{n}p_h \\ 
& > \sum_{j=1}^{v'}r_j+r_{v'+1}\sum_{h=1}^{n}q_h \\
&= \sum_{i=1}^{v'd+n} (\mathbf{q} \otimes \mathbf{r})_i^{\downarrow}
\end{split}
\end{equation}
Where we used the fact that $\sum_{h=1}^{n}p_h > \sum_{h=1}^{n}q_h$ from the definition of $n$ and that $\sum_{h=1}^{d}p_h = \sum_{h=1}^{d}q_h = 1$. From (\ref{res2}) and because $p_1 \leq q_1$ implies $p_1r_1 \leq q_1r_1$,  we conclude $\mathbf{p} \otimes \mathbf{r} \nprec \mathbf{q} \otimes \mathbf{r}$, which contradicts the assumption that $\mathbf{p} \otimes \mathbf{r} \prec \mathbf{q} \otimes \mathbf{r}$. Thus, we have proved that:
\begin{equation}
\label{halfbound2}
\max_{v \in \{1,2,\cdots,k-1\}}\left(\frac{r_v}{r_{v+1}}\right) < \frac{q_{n+1}}{q_d}
\end{equation}
Combining (\ref{halfbound}) and (\ref{halfbound2}) completes the proof of bound~(\ref{bound2}).

We will now prove bound (\ref{eqn5}). Again, we use proof by contradiction. Let:
\begin{equation}
\label{assume3}
\frac{r_1}{r_k} \leq \max_{l\in\mathcal{L}}\left(\frac{q_l}{q_{l+1}}\right) \equiv \frac{q_{l'}}{q_{l'+1}}
\end{equation}
Where $l'$ is the specific $l \in \mathcal{L}$ satisfying the maximum. Let $(\mathbf{p} \otimes \mathbf{r})'$ and $(\mathbf{q} \otimes \mathbf{r})'$ be the product Schmidt vectors when $\mathbf{r}$ is the maximally entangled state. From (\ref{assume3}), we have $q_{l'+1}r_1 \leq q_mr_{l'}$ which implies that the first $kl'$ largest elements of $(\mathbf{q} \otimes \mathbf{r})^{\downarrow}$ are given by $\{q_xr_y\}$ with $1\leq x \leq l'$ and $1\leq y \leq k$. The result is that the first $kl'$ elements of $(\mathbf{q} \otimes \mathbf{r})^\downarrow$ are identical to the case when $\mathbf{r}$ is the maximally entangled state. That is, ${\sum_{i=1}^{kl'}(\mathbf{q} \otimes \mathbf{r})^{\downarrow} = \sum_{i=1}^{kl'}(\mathbf{q} \otimes \mathbf{r})'^{\downarrow}}$. It must follow that:
\begin{equation}
\label{rescase1}
\begin{split}
\sum_{i=1}^{kl'}(\mathbf{p} \otimes \mathbf{r})_{i}^{\downarrow} & \geq \sum_{i=1}^{kl'}(\mathbf{p} \otimes \mathbf{r})_{i}'^{\downarrow}\\
& = \left(\sum_{h=1}^{l'}p_h\right)\left(\sum_{j = 1}^{k}r_j \right) = \sum_{h = 1}^{l'} p_{h}\\
 & > \sum_{h' = 1}^{l'} q_{h'} = \left(\sum_{h'=1}^{l'}q_{h'}\right)\left(\sum_{j' = 1}^{k}r_{j'} \right)\\
 & = \sum_{i=1}^{kl'}(\mathbf{q} \otimes \mathbf{r})_{i}'^{\downarrow} =  \sum_{i=1}^{kl'}(\mathbf{q} \otimes \mathbf{r})_{i}^{\downarrow}
\end{split}
\end{equation}
Where we used the fact that $\sum_{h = 1}^{k}r_h = 1$ and that $\sum_{j=1}^{l'}p_j > \sum_{j=1}^{l'}q_j$ since $l' \in \mathcal{L}$. Notice also that $\sum_{i=1}^{kl'}(\mathbf{p} \otimes \mathbf{r})_{i}^{\downarrow} \geq \sum_{i=1}^{kl'}(\mathbf{p} \otimes \mathbf{r})_{i}'^{\downarrow}$ by definition. From (\ref{rescase1}) and because $p_1 \leq q_1$ implies $p_1r_1 \leq q_1r_1$,  we can conclude $\mathbf{p} \otimes \mathbf{r} \nprec \mathbf{q} \otimes \mathbf{r}$, which contradicts the assumption that $\mathbf{p} \otimes \mathbf{r} \prec \mathbf{q} \otimes \mathbf{r}$. Therefore we have derived:
\begin{equation*}
\frac{r_1}{r_k} > \max_{l\in\mathcal{L}}\left(\frac{q_l}{q_{l+1}}\right)
\end{equation*}
This completes the proof of bound (\ref{eqn5}) and the proof of Theorem 1. \QED
\newline
\newline
\textbf{Corollary:} If $q_1 = q_m$ or $q_{n+1} = q_d$, then there is no Schmidt vector $\mathbf{r}$ that catalyses the transformation.
\newline
\newline
\textbf{Proof:} By definition, $r_{v'} \geq  r_{v'+1}$. If $q_1 = q_m$, then $\frac{r_{v'}}{r_{v'+1}} < 1$ implying $r_{v'} < r_{v'+1}$. This contradicts $r_{v'} \geq r_{v'+1}$. Thus, $q_1 \neq q_m$ if a catalyst is to exist for the transformation. By a similar argument, $q_{n+1} \neq q_d$ if a catalyst is to exist for the transformation. This completes the proof. \QED
\newline
\newline
This corollary is strongest when $d = 4$ because $m,n \neq 1,d-1,d$ which implies that $m = n = 2$. Specifically, it states that if $d=4$ and $q_1 = q_2$ or $q_3 = q_4$, then $\mathbf{r}$ cannot catalyse the transformation.

Bound (\ref{bound2}) limits how similar to the product state the probability distribution of $\mathbf{r}$ can be. If the maximum ratio between two subsequent elements of $\mathbf{r}$ is too large, then $\mathbf{r}$ is not entangled enough to catalyse the transformation. In other words, (\ref{bound2}) represents the minimum entanglement $\mathbf{r}$ must contain in order to facilitate the transformation. Bound (\ref{eqn5}) limits how flat (how close to the maximally entangled state) the probability distribution of $\mathbf{r}$ can be. It states that if the ratio between the first element and last element of $\mathbf{r}$ is too close to one, then $\mathbf{r}$ is too entangled to catalyse the transformation. Thus, this bound represents the maximum entanglement a catalyst vector can contain and still potentially catalyse the transformation. If bounds (\ref{bound2}) and (\ref{eqn5}) are both satisfied, it is not possible to determine with certainty whether or not $\mathbf{p} \otimes \mathbf{r}$ and $\mathbf{q} \otimes \mathbf{r}$ are incomparable. Therefore, bounds (\ref{bound2}) and (\ref{eqn5}) represent the optimal bounds that can be derived using the approach presented in this paper.
\subsection{A Bound on Minimum Catalyst Dimension} 
In this section, we identify a lower bound on the dimension $k$ of an ancillary Schmidt vector $\mathbf{r}$ that catalyses a particular incomparable state transformation. Because this new bound utilizes the results in {Theorem 1}, we will make the following notational simplifications. Let:
\begin{equation}
\label{simplify}
a \equiv \min\left(\frac{q_1}{q_{m}},\frac{q_{n+1}}{q_d}\right) \quad \text{and} \quad b \equiv \max_{l \in \mathcal{L}}\left(\frac{q_{l}}{q_{l+1}}\right)
\end{equation}
With these notations, the bound on catalyst dimension is presented in Theorem 2:
\newline
\newline
\textbf{Theorem 2:}  For any incomparable Schmidt vectors $\mathbf{p}$, $\mathbf{q} \in \mathbb{R}^d$ and for any Schmidt vector $\mathbf{r} \in \mathbb{R}^k$, if $\mathbf{p} \otimes \mathbf{r} \prec \mathbf{q} \otimes \mathbf{r}$ , then the dimension of $\mathbf{r}$ must satisfy:
\begin{equation}
\label{newdimbound}
k > \frac{\ln(b)}{\ln(a)}+1
\end{equation}
\newline
\newline
\textbf{Proof:} We know from Theorem 1 that $r_{v'}/r_{v'+1} < a$, where $v'$ is the specific $v$ satisfying the maximum in bound (\ref{bound2}). Additionally, we know from (\ref{eqn5}) that $r_1/r_k > b$. We want to find the minimum dimension $k$ such that bound (\ref{eqn5}) is satisfied. The ratio $r_1/r_k$ is related to all the ratios of the form $r_v/r_{v+1}$, $v \in \{1,2,\cdots,k-1\}$ as follows:
\begin{equation}
\label{relation}
\frac{r_1}{r_k} = \prod_{v=1}^{k-1} \frac{r_v}{r_{v+1}}
\end{equation}

The minimum $k$ required for bound (\ref{eqn5}) to be satisfied occurs when $r_v/r_{v+1} = r_{v'}/r_{v'+1}$ for all ${v \in \{1,2,\cdots,k-1\}}$. In addition, the maximum value of $r_{v'}/r_{v'+1}$ is $a$. For this reason, we will relax the strict inequality of (\ref{bound2}) for the purposes of deriving the bound on dimension and consider the case in which $r_v/r_{v+1} = r_{v'}/r_{v'+1} = a$ for all $v \in \{1,2,\cdots,k-1\}$. This constraint ensures that we are considering the extremal case in which $k$ is minimized. Using equation (\ref{relation}), we identify that $r_1/r_k = a^{k-1}$ in this case. Using this identity and bound (\ref{eqn5}), we get the following relationship between $a$ and $b$:
\begin{equation}
\frac{r_1}{r_k} = a^{k-1} > b
\end{equation}
Solving this relationship for $k$ yields:
\begin{equation*}
k > \frac{\ln(b)}{\ln(a)}+1
\end{equation*} 
This completes the proof. \QED
\newline
\newline

Because bound (\ref{newdimbound}) is only dependant on $a$ and $b$, it is exclusively determined by the Schmidt coefficients of the target state $\mathbf{q}$. The right hand side of bound (\ref{newdimbound}) is strictly greater than one. The bound is non-trivial when $k >2$, which occurs if and only if $b>a$. A non-trivial example demonstrating the utility of bound (\ref{newdimbound}) is shown in the next section.
\section{\label{regions}Examples}
In this section, the maximum probability of a state transformation and the majorization distance will be harnessed to observe specific examples of ELOCC transformations that emphasize the utility of bounds (\ref{bound2}), (\ref{eqn5}) and (\ref{newdimbound}). Specifically, we begin with a simple example in the qubit catalyst case that shows bounds (\ref{bound2}) and (\ref{eqn5}) limiting the set of potential catalysts. We then prove the existence of ELOCC state transformations with more than one distinct region of potential qubit catalysts fascillitating them. This indicates a difficulty of deriving further bounds on ELOCC transformations. Finally, a higher dimentional (non-qubit) transformation will be presented providing a comparison of the bounds shown in \cite{Sanders2009} with the three bounds derived in this paper.
\subsection{Maximum Probability of Transformation and Majorization Distance}
For any two incomparable bipartite states $|\psi\rangle$ and $|\phi\rangle$, there exists a maximum probability ($<1$) that the transformation  $|\psi\rangle \rightarrow |\phi\rangle$ is achieved. The maximum probability of a pure bipartite state transformation was first proved by \cite{Vidal1999}. We wish to observe how the maximum probability of an incomparable transformation varies when a qubit Schmidt vector $\mathbf{r} = (1-t,t), \ t \ \in [0,1/2]$ is borrowed to catalyse the transformation. The modified maximum probability of the product state transformation is:
\begin{equation}
\label{prob}
P_{max}(|\psi\rangle \otimes |\chi\rangle \rightarrow |\phi\rangle \otimes |\chi\rangle) = \min_{l^* \in \{1,2,\cdots,2d\}}\frac{E_{l^*}(|\psi\rangle \otimes |\chi\rangle)}{E_{l^*}(|\phi\rangle \otimes |\chi\rangle)}
\end{equation}
where $E_{l^*}(|\psi\rangle \otimes |\chi\rangle)=1-\sum_{i = 1}^{l^*-1} (\mathbf{p}\otimes\mathbf{r})_{i}^{\downarrow}$. When $l^* = 1$, both sums in (\ref{prob}) are zero, making $P_{max}(t) = 1$. Thus, the maximum probability of a transformation never exceeds one.

In addition to the maximum probability, the majorization distance provides a measure of how close a vector $\mathbf{q}$ is to majorizing another vector $\mathbf{p}$ and was first defined by \cite{Horodecki2017} in the context of approximate majorization. If $\mathbf{r}$ is a qubit Schmidt vector, then analogously to the maximum probability, we can observe how the majorization distance varies when $\mathbf{r}$ is borrowed to catalyse the transformation. The modified majorization distance of the product state transformation is:
\begin{equation}
\label{distance}
\delta(t) = 2 \max_{l^* \in \{1,\cdots,2d\}}\sum_{i=1}^{l^*}((\mathbf{p} \otimes \mathbf{r})_{i}^{\downarrow}-(\mathbf{q} \otimes \mathbf{r})_{i}^{\downarrow})
\end{equation}

Since both (\ref{prob}) and (\ref{distance}) are functions of $t$, we can plot $P_{max}(t)$ and $\delta(t)$ against $t$ for $t \in [0,1/2]$. Both the maximum probability and the majorization distance reveal which values of $t$ make the Schmidt vector $\mathbf{r}$ an effective catalyst for the transformation. Specifically, when $P_{max}(t) = 1$, or alternatively, when $\delta(t) = 0$, $\mathbf{r}$ is an effective catalyst for the transformation. Both of these functions will be used to visualize specific examples in subsequent sections.
\subsection{A Simple Qubit Example}
In the qubit catalyst case, bounds (\ref{bound2}) and (\ref{eqn5}) produce direct upper and lower bounds on potential ancillary states because $r_{v'}/r_{v'+1} = r_1/r_2 = r_1/r_k$.  Specifically, if it is assumed that $\mathbf{r} = (1-t,\, t)$, where $t \in [0,1/2]$, then:
\begin{equation}
\label{2db1}
 \frac{1}{a+1} < t < \frac{1}{b+1}
\end{equation}
Where $a$ and $b$ are the right hand side of (\ref{bound2}) and (\ref{eqn5}) respectively defined in equation (\ref{simplify}). As a simple example, consider the vectors:
\begin{equation}
\begin{split}
\mathbf{p} & =(0.45,\, 0.35,\, 0.12,\, 0.08) \\
\mathbf{q} & = (0.56,\, 0.21,\, 0.17,\, 0.06)
\end{split}
\end{equation}

\begin{figure}[h!] 
\makebox[\linewidth]{
\includegraphics[width=\columnwidth]{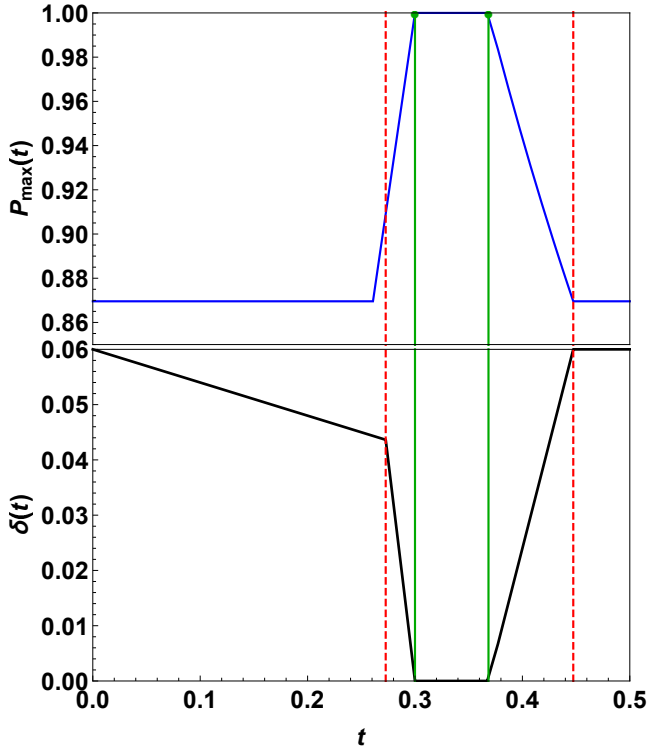}
	}
	\caption{\label{fig1} (color online) The maximum probability of transformation (top) and the majorization distance (bottom) plotted against $t$. The region in which $P_{max}(t) = 1$ and $\delta(t) = 0$ is the region of catalysis. The vertical solid green lines bound the exact region of qubit catalysis for this transformation. The left and right vertical red dashed lines represent the bounds on minimum and maximum catalyst entanglement, respectively.}
\end{figure} 
Clearly these vectors are incomparable since $0.45 < 0.56$ but $0.45 + 0.35 > 0.56 + 0.21$.  From (\ref{2db1}), we see that $0.272727 < t < 0.447368$. This region of potential catalysis can be visualized by plotting the maximum probability of the transformation and the majorization distance in the form of (\ref{prob}) and (\ref{distance}) against $t$. This is shown in FIG. \ref{fig1}.

FIG. \ref{fig1} shows that (\ref{2db1}) is a good approximation of the region of catalysis but does not represent strict minimum and maximum bounds. The bound on minimum entanglement is shown as the left vertical red dashed line while the bound on maximum entanglement  is shown as the right vertical red dashed line. From the solid vertical green lines, It is clear that both the maximum probability of the transformation and the majorization distance convey the same information on the exact region of catalysis. Namely, the range of $t$ to which $\delta(t) = 0$ exactly matches the range of $t$ in which $P_{max}(t) = 1$. 

In FIG. \ref{fig1}, the bounds on catalyst entanglement appear to overlap with specific points to which the qualitative behaviour of $P_{max}(t)$ and $\delta(t)$ change. The reason for this is that bounds (\ref{bound2}) and (\ref{eqn5}) were derived by analysing specific elements of $(\mathbf{p} \otimes \mathbf{r})^{\downarrow}$ and $(\mathbf{q} \otimes \mathbf{r})^{\downarrow}$. At the precise values of $t$ where either bound (\ref{bound2}) or (\ref{eqn5}) become satisfied, two elements of $(\mathbf{q} \otimes \mathbf{r})^{\downarrow}$ change place (for example $q_{l'+1}r_1 > q_{l'}r_k$ becomes  $q_{l'+1}r_1 \leq q_{l'}r_k$). This swap in elements often causes the maximum in (\ref{prob}) and/or (\ref{distance}) to be attained by a different value of $l^*$, which ultimately changes the qualitative behaviour of the plots.
\subsection{Multiple Regions of Catalysis}
In this section, the existence of a new class of catalytic state transformations is proven. Namely, the existence of incomparable vectors with multiple distinct ranges of potential qubit catalysts is shown. Consider the intervals $I_1 = [w,x]$, $I_2 = (x,y)$ and $I_3 = [y,z]$, where $0 \leq w < x < y < z \leq 1/2$. It will be shown that there exists $w$, $x$, $y$, $z$ and incomparable Schmidt vectors $\mathbf{p}$ and $\mathbf{q}$ such that when $t \in I_1$ or $t \in I_3$, $\mathbf{r} = (1-t,t)$ is an effective catalyst for the state transformation, while when $t \in I_2$, $\mathbf{r}$ is not an effective catalyst for the transformation. That is, there are two distinct regions of qubit catalysts that will do the job. 
\begin{figure}[h!] 
\makebox[\linewidth]{
\includegraphics[width=\columnwidth]{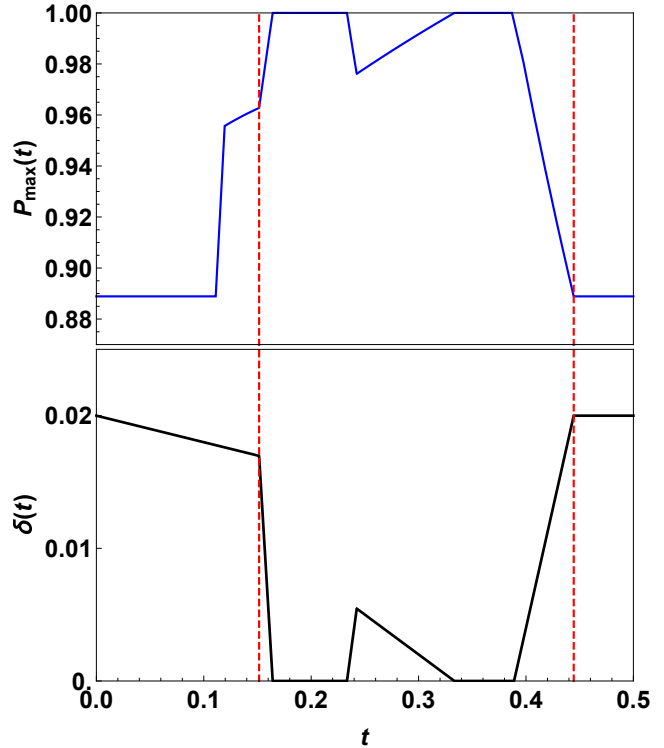}
	}
	\caption{\label{fig4} (color online) The maximum probability of the state transformation (top) and the majorization distance (bottom) as functions of $t$. In this case, there are two ranges of $t$ to which $P_{max} = 1$ and $\delta(t) = 0$, leaving a range in between where $\mathbf{r}$ is not an effective catalyst for the transformation.  Again, the left and right vertical red dashed lines represent the bounds on minimum and maximum catalyst entanglement respectively.}
\end{figure}

For example, let
\begin{equation}
\begin{split}
\label{example} 
\mathbf{p} & = (0.49,\, 0.30,\, 0.13,\, 0.06,\, 0.02)\\
\mathbf{q} & =  (0.56,\, 0.25,\, 0.10,\, 0.08,\, 0.01)
\end{split}
\end{equation}
 The bounds on qubit catalyst entanglement (\ref{2db1}) state that $0.151515 < t < 0.444444$. FIG. \ref{fig4} shows the maximum probability of the transformation and the majorization distance as functions of $t$ for this example. It is remarkable to see that when $t = 0.2$ and when $t = 0.35$, $\mathbf{r}$ effectively catalyses the state transformation, however when $t = 0.3$ it does not! 

The discovery of this class of incomparable state transformations has a few important implications. Namely, the existence of multiple regions of catalysis makes it much harder to find exact bounds on qubit ELOCC transformations. No matter how close the bounds given by (\ref{2db1}) are to the minimal or maximal effective catalyst, it is not guaranteed that all $\mathbf{r}$ that lie within the bounded region catalyse the transformation. There may be additional gaps (such as in FIG. \ref{fig4}) where catalysis does not occur. In order to fully characterize ELOCC transformations even in the qubit case, one would have to find conditions on $\mathbf{p}$ and $\mathbf{q}$ to which multiple regions of catalysis exist. More specifically, one would have to find bounds that accurately disregard incomparable regions that are bounded by effective catalysts. Nevertheless, the bounds presented in this paper substantially restrict the domain of catalysts that need to be considered for any given transformation.

In general, it appears that there are no incomparable four-dimensional vectors $\mathbf{p}$ and $\mathbf{q}$ that have multiple regions of qubit catalysis as in the example shown in FIG. \ref{fig4}. Moreover, state transformations are not limited to only two regions of catalysis. Examples have been found where there are three or more regions of potential catalysts. It is conjectured that there is no limit on the maximum number of distinct regions of potential qubit catalysts allowing a particular ELOCC transformation. Furthermore, the likelihood of multiple regions appears to rise as the dimension of the Schmidt vectors $\mathbf{p}$ and $\mathbf{q}$ becomes large. The most intuitive explanation as to why this is the case is that larger dimensional Schmidt vectors tend to have a larger number of element swaps in $(\mathbf{p} \otimes \mathbf{r})^{\downarrow}$ and $(\mathbf{q} \otimes \mathbf{r})^{\downarrow}$ as $t$ varies from zero to one half. Thus, there is more likelihood that for certain $t$ values, the product vectors may become incomparable after a period of effective catalysis. This would also explain why four-dimensional Schmidt vectors rarely, if at all, exhibit disjoint regions of qubit catalysis. In this case, there is simply not enough element swaps for disjoint regions of catalysis to occur. In all, it appears that classifying the set of transformations with disjoint regions of catalysis is a significant challenge because two different incomparable state transformations with only a slight deviation in Schmidt coefficients can have vastly different catalytic properties. For this reason, we leave the classification of disjoint ELOCC transformations for future work.

\subsection{\label{compare} A Higher Dimensional Example}
We will now show a higher dimensional (non-qubit) catalytic example demonstrating that (\ref{bound2}), (\ref{eqn5}) and (\ref{newdimbound}) are able to impose conditions on incomparable state transformations that previous bounds, first discovered in \cite{Sanders2009}, can not. This example will demonstrate the breadth of information that bounds (\ref{bound2}), (\ref{eqn5}) and (\ref{newdimbound}) can provide for a particular non-qubit transformation.

The bounds in \cite{Sanders2009} were dicovered using the class of Schur-convex/concave functions. Any real valued function $f$ is said to be Schur-convex if $\mathbf{p} \prec \mathbf{q}$ implies $f(\mathbf{p}) \leq f(\mathbf{q})$. The function $f$ is said to be Schur-concave if $-f$ is Schur-convex \cite{Diggle2010}.

By exploiting the Schur convexity/concavity of the elementary and power sum symmetric polynomials using Newton's identities \cite{E.W.Weisstein}, a bound on the minimum dimension $\mathbf{r}$ must be in order to catalyse a particular incomparable transformation was presented in \cite{Sanders2009}. Namely it was found that:
\begin{equation}
\label{dimbound}
k \geq \frac{\log_2(e_{d-1}(\mathbf{q}))-\log_2(e_{d-1}(\mathbf{p}))}{\log_2(e_{d}(\mathbf{p}))-\log_2(e_{d}(\mathbf{q}))} + 1
\end{equation}
Where $e_j(\mathbf{p}), \ j \in \{0,1 \cdots d\}$ is the $j$th elementary symmetric polynomial of $\mathbf{p}$. (\ref{dimbound}) is only non-trivial when the logarithmic term is larger than one (when $k \geq 2$). In addition to (\ref{dimbound}), the following bound was also presented in \cite{Sanders2009}:
\begin{equation}
\mathcal{R}(\mathbf{r}) \geq -\frac{e_3(\mathbf{p})-e_3(\mathbf{q})}{e_2(\mathbf{p})-e_2(\mathbf{q})}
\end{equation}
Where 
\begin{equation}
\label{valbound}
\mathcal{R}(\mathbf{r}) = \frac{e_2(\mathbf{r})-2e_3(\mathbf{r})}{1-2e_2(\mathbf{r})+3e_3(\mathbf{r})} \geq 0
\end{equation}
Note that $e_2(\mathbf{p})-e_2(\mathbf{q}) > 0$ is a monotone under ELOCC, however $e_3(\mathbf{p})-e_3(\mathbf{q})$ is not \cite{Sanders2009}. Therefore, in order for (\ref{valbound}) to be non-trivial, $e_3(\mathbf{p})-e_3(\mathbf{q}) < 0$. Because both (\ref{dimbound}) and (\ref{valbound}) must satisfy conditions to be non-trivial, they only provide information on the catalyst Schmidt vector $\mathbf{r}$ for a subset of all catalytic examples. 

To demonstrate this, consider the vectors:
\begin{equation}
\begin{split}
    \mathbf{p} & = (0.47,\, 0.38,\, 0.13,\, 0.02) \\
    \mathbf{q} & = (0.53,\, 0.31, \,0.15,\, 0.01)
\end{split}
\end{equation}
Bound (\ref{dimbound}) states that $k \geq 0.918917$ which rounds to $k \geq 1$. Furthermore, bound (\ref{valbound}) states that $\mathcal{R}(\mathbf{r}) \geq -0.170824$. Because at minimum, $k \geq 2$ and $\mathcal{R}(\mathbf{r})\geq 0 \ \forall \ \mathbf{r}$, these two bounds are trivial for this example.

On the contrary, bound (\ref{bound2}) states that ${r_{v'}/r_{v'+1} < 1.70968}$ and bound (\ref{eqn5}) states that $r_1/r_k > 2.06667$. Thus, the bounds on minimum and maximum entanglement produce non-trivial results for this incomparable transformation. Bound (\ref{newdimbound}) reveals that $k>2.35359$ which rounds to $k \geq 3$. Therefore, we have successfully identified that no qubit catalyst exists for this transformation. This analysis shows that the new bounds presented in this paper provide conditions on the subset of incomparable state transformations that could not be classified with the bounds presented in \cite{Sanders2009} alone. To conclude, the set of all catalytic transformations to which one can identify meaningful information regarding potential catalysts has been greatly increased.
\section{\label{conc}Conclusions}
In this paper, we address the following problem. Consider a state transformation $|\psi \rangle \rightarrow |\phi \rangle$ that cannot be achieved with LOCC alone. What catalyst $|\chi \rangle$ changes the process into the ELOCC transformation ${|\psi \rangle \otimes |\chi \rangle \rightarrow |\phi \rangle \otimes |\chi \rangle}$? We provide a partial answer to this question by considering the minimum deviation a potential catalyst must have from both the product state and the maximally entangled state for an ELOCC transformation to become possible. In particular, we have shown that for any incomparable Schmidt vectors $\mathbf{p}$ and $\mathbf{q}$, if the Schmidt vector $\mathbf{r}$ is a catalyst for the transformation, then it must have enough entanglement to satisfy (\ref{bound2}), but not so much entanglement that it violates (\ref{eqn5}). We have also identified a bound restricting the minimum dimension of $\mathbf{r}$. This lower bound depends only on the Schmidt coefficients of the target state $\mathbf{q}$ and is non-trivial for a large set of catalytic examples.

The solution in this paper is only partial as bounds (\ref{bound2}) and (\ref{eqn5}) are not tight bounds excluding all catalysts that are not effective. Nor is the bound on catalyst dimension a tight bound on the true minimum dimension for all catalytic examples. However, these bounds substantially restrict the set of potential catalysts for any particular state transformation by providing restrictions on the amount of entanglement a catalyst state may contain. We have shown an example demonstrating that bounds (\ref{bound2}), (\ref{eqn5}) and (\ref{newdimbound}) provide conditions on the set of catalytic transformations which the previous bounds in \cite{Sanders2009} could not. Of course, in an ideal setting, both the bounds in \citep{Sanders2009} and the ones presented in this paper will be used in conjunction to provide the best possible restrictions on potential catalysts. 

The bounds presented in this paper have vast importance in entanglement theory, through the application of ELOCC transformations in quantum information processes such as teleportation, where one may desire a particular state for maximum efficiency. Furthermore, these bounds have implications in quantum thermodynamics, where athermality is the core resource rather than entanglement. In terms of athermality, the most resourceful state is the product state and the least resourceful state is the maximally entangled state. In the case of thermomajorization, the initial state must majorize the final state for the transformation to be realized with certainty  \cite{Gouragain}. Because of these parallels between quantum thermodynamics and entanglement theory, the bounds on minimum and maximum entanglement are equivalently the bounds on maximum and minimum athermality respectively. The bound on catalyst dimension also remains integral in the thermodynamic regime, representing the minimum dimension of a potential catalyst heat engine undergoing a cyclic process in a thermal operation. In all, the bounds presented in this paper have scope that extends beyond entanglement theory to any other resource theory that utilizes majorization as the necessary and sufficient condition for state transformations. It has been shown that when a separable operation acts on a pure bipartite state, it is governed by a majorization condition that is identical to the particular case of LOCC \cite{Vlad}. Consequently, the results presented in this paper extend beyond ELOCC transformations to the larger, more general set of ancillary assisted separable operations.

Additionally, we have shown the existence of qubit ELOCC transformations that have multiple distinct regions of effective qubit catalysts allowing them. The existence of these examples demonstrates how difficult exact bounds on ELOCC transformations would be to achieve, even in the qubit case. Due to the similarities between the resource theories of entanglement and athermality, the existence of disjoint regions of catalysis poses an equally important problem for bounding catalyst heat engines in quantum thermodynamic processes. To proceed from this work to a more general description of ELOCC transformations, one would have to determine the conditions for when an incomparable state transformation has more than one continuous region of catalysis and furthermore would have to bound these additional regions precisely. It is vitally important to fully understand transformations that require the use of ancillary systems as they provide additional conversion power with little to no increase in resource cost. This conversion power is valuable as most quantum information processes require a precise state for optimal efficiency. This work provides a crucial step in achieving a full understanding of ELOCC state transformations.

\section*{Acknowledgements}
The authors would like to thank John Burniston for useful discussions related to the topic of this paper and for vital feedback and editing. The authors acknowledge support from the Natural Sciences and Engineering Research Council of Canada (NSERC). This work was completed as a part of an undergraduate physics thesis project (PHYS 598) at the University of Calgary.

\bibliography{phys598finalupdate}
\end{document}